\voffset=.3in
\magnification=\magstep1
\parskip=4pt plus 2pt minus 2pt 
%

\centerline{\bf FATHER TIME. II. A PHYSICAL BASIS BEHIND FEYNMAN'S IDEA OF} 
\centerline {\bf ANTIPARTICLES MOVING BACKWARD IN TIME, AND AN EXTENSION}
\centerline{\bf OF THE CPT THEOREM TO INCLUDE NON-LOCAL GAUGE FIELDS}
\vskip 2pc
\centerline{T K Rai Dastidar\footnote *{Atomic \& Molecular Physics Group,  
Department of Materials Science. 
Electronic address : mstkrd@mahendra.iacs.res.in} and Krishna Rai
Dastidar\footnote {\dag}{Department of Spectroscopy. Electronic address :
spkrd@mahendra.iacs.res.in}}
\vskip 1pc
\centerline{\it Indian Association for the Cultivation of Science, Calcutta 
700\thinspace 032, India}
\vskip 3pc
\centerline {\bf Abstract}
\vskip 1pc
It was demonstrated in a recent paper (Mod.Phys.Lett. A{\bf 13}, 1265 
(1998) ; hep-th/9902020) that the existence of a non-thermodynamic arrow 
of time at the atomic level is a fundamental requirement for conservation 
of energy in matter-radiation interaction. Since the universe consists of 
two things only --- energy and massive matter --- we argue that as a 
consequence of this earlier result, particles and antiparticles must 
necessarily move in opposite directions in time. Our result further
indicates that the CPT theorem can be extended to cover non-local 
gauge fields.
\vfill\eject
\noindent {\bf I. Introduction}
\vskip 1pc
Some time ago [1] we obtained a result that matter fields possess a 
non-local U(1) gauge transformation symmetry that necessitates invoking
a non-local electromagnetic field, and that a non-thermodynamic arrow
of time at the quantum (atomic) level is enforced upon us as a necessary
condition for conservation of energy in matter-radiation interactions. Based
upon this result, it was subsequently predicted [2] that the spectrum of a 
blackbody radiation passing through gaseous matter would show an apparent 
deviation from the Planck formula. A very simple experimental test for 
verifying this prediction has been suggested in [2], where it was also shown 
that the results of a balloon-borne measurement [3] of the cosmic microwave 
background radiation (CMBR) spectrum does seem to show such a 
feature\footnote{$^{\rm a}$}{In [2] we also discuss why the cosmic background 
explorer ({\it COBE\/}) satellite measurements of the CMBR spectrum [4] 
{\it do not show\/} this feature.}. 

Subject to verification of the above theoretical prediction, we argued
recently [5] that, as a necessary condition for energy conservation in
matter-radiation interactions, all particles in the universe subject to
electromagnetic interactions should follow this one and the same arrow of 
time through the agency of the CMBR\footnote{$^{\rm b}$}{The restriction, 
that the particles must be subject to electromagnetic interactions, simply 
reflects the present status of the theory (which is undergoing continual 
development), and there seems to be no reason why it should be fundamental~; 
work is in progress to see if it can be removed.}. The present paper, which 
is a companion to [5], addresses reactions which do not conserve energy, but 
involve mass$\leftrightarrow$energy conversion, e.g. particle-antiparticle
annihilation, pair production, particle decay etc.
\vskip 1pc
\noindent {\bf II. Theory and Results}
\vskip 1pc
To make the present paper self-contained, we briefly recapitulate here
how the time's arrow appears in the theory as a necessary condition for 
conservation of energy in the atomic level. (See [1] for details.) A matrix 
element for matter-radiation interaction with the non-local electromagnetic 
field potential ${\cal A}(x,x')$ where {\it x}$\equiv$({\bf r},{\it t}) is 
typically of the form
$$M_{fi}(t)\propto\langle\psi_f(x),n_f|{\hat p}.
{\cal A}(x,x')|\psi_i(x'),n_i\rangle,$$
the integration running over $d^3r$ and $d^4x'$. 
Special relativity requires that the time integral over $dt'$ should run 
from $-\infty$ to $(t-\rho/c)$ [where $\rho=|${\bf r} -- {\bf r}$'|$] 
for the retarded interaction, and from
$(t+\rho/c)$ to $+\infty$ for the advanced interaction. After Fourier
expanding ${\cal A}$ in two sets of photon modes (see, e.g. [6]), the
time integration over $dt$ gives us the energy-conserving
delta-function $\delta(E_f,E_i+\hbar\omega+\hbar\omega')$ in the 
correlated two-photon absorption process (see below ; this is one of the new 
results in [1]). Formally, this time integration runs from $t=-\infty$ to 
$t=+\infty$. However, if we agree to define that the atom has been excited,
i.e. the matrix element has come into existence, at the time $t=0$,
then it is at once obvious that the segment $\int_{-\infty}^0 dt$
does not contribute anything to the {\it t}-integral, giving finally a
double integral to be chosen from
$$\int_0^\infty f(t)dt \left (\int_{-\infty}^{t-\rho/c}g(t')dt',\quad
\int_{t+\rho/c}^\infty g(t')dt'\right )$$
where $f(t)=\exp\left [{i\over \hbar}(E_ft-\hbar\omega t)\right ], 
g(t')=\exp\left [-{i\over \hbar}(E_it'+\hbar\omega't')\right ]$.
It is obvious that the retarded interaction gives the
required $\delta$-function for energy conservation, while the
advanced interaction fails to do so. Thus we are constrained to limit 
the non-local temporal correlation {\it to the past only\/}~; 
the fundamental principle of energy conservation
has given us an arrow of time, i.e. the principle of causality, as
a corollary.\footnote{$^{\rm c}$}{As shown in [1], however, we have 
to pay a price for it~;
the correlation with the ``past'' is to be understood in an EPR-like
sense and not in a Lorentz invariant sense, if we are to maintain
energy conservation. (This serves to remind us that quantum mechanics and
special relativity have always been strange bedfellows.) It is, of course, 
just a convention of language that we visualise particles, i.e. the
constituents of our own universe as moving ``forward'' in time, thus 
establishing a causal link with the ``past'' and not with the ``future''.} 

We are thus led to the result that in atomic-scale reactions, forward
motion in time of all the reactants/products is a necessary condition
for conservation of energy. What happens if energy is not conserved~? 
The universe contains but two things --- energy and massive matter ---
and mass$\leftrightarrow$energy conversion is the only possible outcome
of such reactions. However, according to our earlier result, such
``non-conservation'' of energy in atomic-scale reactions can take place 
if and only if one or more of the reactants/products do not move forward in 
time, i.e. move backward in time (since time is one-dimensional).

We may ask the question from the opposite viewpoint : what happens in
a reaction if one or more of the reactants/products travel backward in
time ? According to [1], the answer is simple~; energy will not be 
conserved, hence mass$\leftrightarrow$energy conversion must occur
in such reactions. (Of course, the precise time interval between the
``start'' of the reaction and the mass$\leftrightarrow$energy conversion
depends on the particular characteristics in the process involved~; for 
example, positron-electron annihilation can take place either 
instantaneously or after positronium formation, depending on the 
collision geometry.)

We see therefore that from our time's arrow, we have obtained the result
that in reactions where mass$\leftrightarrow$energy conversion takes
place, at least one of the reactant/product particles must be travelling
in time in a direction opposite to the others. When we recall that
such reactions must involve both ``particles'' and ``antiparticles'', 
then from known properties of antiparticles, we conclude that an antiparticle
can be obtained only under a combined operation of charge conjugation C,
parity P and time reversal T (in any order) on a particle~; similarly, CPT 
operating on an antiparticle gives us a particle. Since the {\it total 
number\/} of particles and antiparticles in any process involving particles 
and antiparticles is always even, the matrix element for the process remains 
invariant under the CPT operation.
(This invariance is not quite self-evident, as we remember that the CPT 
theorem is supposed to be valid for {\it local\/} field theories only~; 
our work seems to have broadened its scope.)

To conclude, we have found a physical basis for Feynman's idea that 
antiparticles can be considered as particles moving backward in time, 
and we have also found that the scope of the CPT theorem can be extended
to cover non-local gauge fields.
\vskip 2pc
We thank Dr J Chakravarti, Department of Theoretical Physics for his 
interest in the ongoing work.
\vfill\eject
\centerline {\bf References}
\vskip 1pc
\item{1.} T K Rai Dastidar and Krishna Rai Dastidar, Mod. Phys. 
Lett. A{\bf 13}, 1265 (1998) ; {\it Errata} A{\bf 13}, 2247 (1998) ;
hep-th/9902020
\item{2.} T K Rai Dastidar, Mod.\thinspace Phys.\thinspace Lett. A
{\bf 14}, 1193 (1999) ; quant-ph/9903043
\item{3.} D P Woody and P L Richards, Phys. Rev. Lett. {\bf 42}, 925
(1979)
\item{4.} J C Mather et al, Astrophys. J. {\bf 420}, 439 (1994)
\item{5.} T K Rai Dastidar, ``Father Time. I. Does the Cosmic Microwave
Background Radiation Provide a Universal Arrow of Time~?'' quant-ph/9903053
(To appear in Mod. Phys. Lett. A).
\item{6.} I M Gel'fand and G E Shilov, {\it Generalised Functions}
(Acad. Press, 1964) Vol. 1, Chap. 2, eq. 1.3(1)
\end